\documentclass[12pt]{article}
\usepackage{amsmath}
\usepackage{amssymb}
\usepackage{bm}
\usepackage{graphicx}
\usepackage[tiny,bf]{titlesec}
\usepackage{hyperref}
\usepackage[backend=biber,style=numeric,citestyle=numeric,backref=true,sorting=none,natbib=true,autopunct=true,uniquename=false,uniquelist=false,giveninits]{biblatex}
\DeclareNameAlias{sortname}{last-first}
\usepackage{tikz}
\usepackage[compat=1.1.0]{tikz-feynhand}
\usepackage{titlesec}
\usepackage{setspace}

\titlespacing\section{0pt}{2ex plus 1ex minus 1ex}{1ex plus 1ex minus 1ex}

\newcommand{\xy}[1]{{\textcolor[hsb]{0,0,.4}{#1}}}
\newcommand{\xz}[1]{{\textcolor[hsb]{.04,.7,.7}{#1}}}
\newcommand{\xr}[1]{{\textcolor[hsb]{0,1,1}{#1}}}
\newcommand{\xg}[1]{{\textcolor[hsb]{.333333,1,1}{#1}}}
\newcommand{\xb}[1]{{\textcolor[hsb]{.666667,1,1}{#1}}}

\newcommand{\yu}{\xy{\uparrow}}
\newcommand{\zu}{\xz{\uparrow}}
\newcommand{\ru}{\xr{\uparrow}}
\newcommand{\gu}{\xg{\uparrow}}
\newcommand{\bu}{\xb{\uparrow}}

\newcommand{\yd}{\xy{\downarrow}}
\newcommand{\zd}{\xz{\downarrow}}
\newcommand{\rd}{\xr{\downarrow}}
\newcommand{\gd}{\xg{\downarrow}}
\newcommand{\bd}{\xb{\downarrow}}

\newcommand{\im}{i}
\newcommand{\ee}{e}
\newcommand{\unit}[1]{\, {\rm #1}}

\newcommand{\Upup}{{{\Uparrow}{\uparrow}}}
\newcommand{\Downdown}{{{\Downarrow}{\downarrow}}}
\newcommand{\Downup}{{{\Downarrow}{\uparrow}}}
\newcommand{\Updown}{{{\Uparrow}{\downarrow}}}

\newcommand{\Lchiral}{L}
\newcommand{\Rchiral}{R}

\newcommand{\Spin}{{\rm Spin}}
\newcommand{\SU}{{\rm SU}}
\newcommand{\U}{{\rm U}}

\newcommand{\bgamma}{\bm{\gamma \mkern-11.5mu \gamma}}
\newcommand{\bH}{\bm{H}}
\newcommand{\bT}{\bm{T}}

\begin{document}

\pagenumbering{gobble}
\hyphenpenalty=3000

\newcommand{\spinelevenonechartfig}{
    \begin{figure*}[t!]
    \begin{center}
    \leavevmode
    \includegraphics[scale=.86]{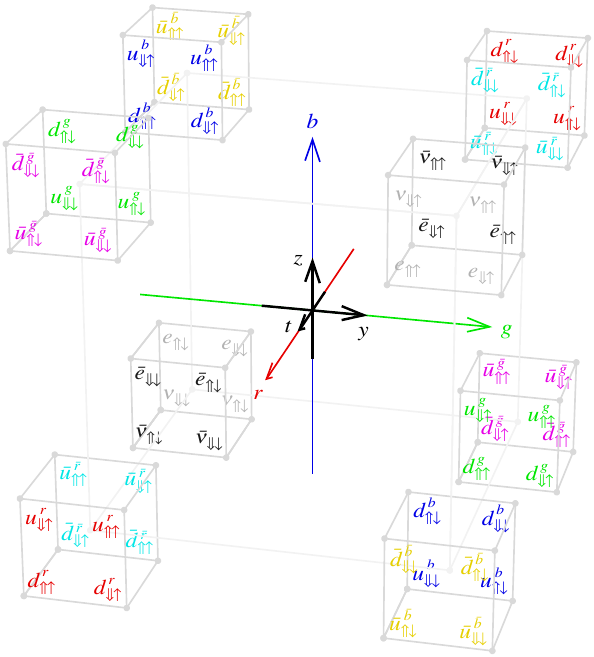}
    \caption[Chart of Spin(10) fermions]{
    \label{spin111chart}
A generation (the electron generation) of 64 fermions
arranged according to their $\Spin(11,1)$ $tyzrgb$ charges.
The eight boxes are distinguished by their color $rgb$ bits.
The fermions in each of the eight boxes
are distinguished by their time and weak $tyz$ bits.
Flipping the $t$-bit of a fermion flips the fermion to its antifermionic partner
of opposite boost and the same spin.
Flipping all 6 $tyzrgb$ bits of a fermion flips its Dirac boost and spin bits,
thereby transforming the fermion to its Weyl companion.
    }
    \end{center}
    \end{figure*}
}


\markboth{A. J. S. Hamilton}{Six Bits}

\title{\bf\large Six Bits}

\author{\normalsize
A. J. S. Hamilton \\
\small
JILA, Box 440, U. Colorado, Boulder, CO 80309, USA \\
\small
Andrew.Hamilton@colorado.edu
}

\date{\small\today}

\maketitle

\begin{abstract}
\noindent
The spinors of the group $\Spin(N)$ of rotations in $N$ spacetime dimensions
are indexed by a bitcode with $[N/2]$ bits.
A well-known promising grand unified group that contains
the standard-model group is $\Spin(10)$.
Fermions in the standard model
are described by five bits $yzrgb$,
consisting of two weak bits $y$ and $z$,
and three color bits $r$, $g$, $b$.
If a sixth bit $t$ is added, necessary to accommodate a time dimension,
then the enlarged $\Spin(11,1)$ geometric algebra contains the
standard model and Dirac algebras as commuting subalgebras,
unifying the four forces of Nature.
There is a unique minimal symmetry-breaking chain
and associated multiplet of Higgs fields
that breaks $\Spin(11,1)$ to the standard model.
Unification to the Pati-Salam group
$\Spin(4)_w {\times} \Spin(6)_c$ is predicted at $10^{12} \unit{GeV}$,
and grand unification at $10^{15} \unit{GeV}$.
The grand Higgs field breaks $t$-symmetry, can drive cosmological inflation,
and generates a large Majorana mass for the right-handed neutrino
by flipping its $t$-bit.
The electroweak Higgs field breaks $y$-symmetry,
and generates masses for fermions by flipping their $y$-bit.
\end{abstract}

\begin{center}
\it
Essay written for the Gravity Research Foundation 2023 Awards
for Essays on Gravitation.
\end{center}

\pagenumbering{arabic}

\section*{Spinors in the standard model of physics}

\noindent
If there is a theory of everything, there is a good chance that spinors
are at the heart of it.
Look around.
All known matter (fermions) is made of spinors.
All known forces arise from symmetries of spinors.

Introduced by Cartan \cite{Cartan:1913,Cartan:1938}
more than a hundred years ago,
spinors, objects of spin $\tfrac{1}{2}$,
constitute the fundamental representation of the group $\Spin(N)$
of rotations in $N$ spacetime dimensions.
Spinors have the intriguing property that their index is a bitcode,
with $[N/2]$ bits in $N$ spacetime dimensions.
The halving of dimensions is associated with the
fact that spinors have a natural complex structure.
%
Associated with each
bit
is a pair of orthonormal basis vectors $\bgamma_k^+$ and $\bgamma_k^-$
in an $N$-dimensional geometric algebra (Clifford algebra)
\cite{Clifford:1878,Hestenes:1966,Hestenes:1987}.
Spinors with $k$ spin up ($\uparrow$) and down ($\downarrow$)
transform with opposite phases $\ee^{\mp \im \theta/2}$
(or opposite boosts $\ee^{\pm \theta/2}$)
under rotations in the 2-dimensional $\bgamma_k^+ \bgamma_k^-$~plane.
The orthonormal vectors $\bgamma^+_k$ and $\bgamma^-_k$
can be interpreted as the real and imaginary parts of a complex vector.

In the 3+1 dimensions of familiar spacetime,
the number of bits is two, a Dirac spinor,
with $2^2 = 4$ complex components.
A Dirac spinor has a spin bit ($\uparrow$ or $\downarrow$)
and a boost bit ($\Uparrow$ or $\Downarrow$).
The Dirac spinor is called right-handed if the spin and boost bits are aligned,
left-handed if anti-aligned.
The geometric algebra in 3+1 dimensions is the Dirac algebra
of Dirac $\gamma$-matrices.

The group $\Spin(10)$ of rotations in 10 dimensions,
proposed in the 1970s by
\cite{Georgi:1975,Fritzsch:1975},
has remained a compelling candidate for
a grand unified group that contains the standard-model group,
the product
$\U(1)_Y {\times} \SU(2)_\Lchiral {\times} \SU(3)_c$
of hypercharge, weak, and color groups.
%
The standard model has 5 conserved charges consisting of
hypercharge $Y$, weak isospin $I_\Lchiral$,
and three colors $R$, $G$, and $B$.
As first pointed out by
\cite{Wilczek:1998},
and reviewed by \cite{Baez:2009dj},
$\Spin(10)$ describes a generation of fermions of the standard model
with a bitcode with $[10/2] = 5$ bits $y , z , r , g , b$
consisting of two weak bits $y$ and $z$,
and three color bits $r , g , b$
(the naming of bits follows \cite{Hamilton:2022b}).
Each bit can be either up or down,
signifying a charge of $+\tfrac{1}{2}$ or $-\tfrac{1}{2}$.
The relation between standard-model charges and $\Spin(10)$ charges is
\begin{equation}
\label{YIc}
  Y
  =
  y + z - \tfrac{2}{3} ( r + g + b )
  \ , \quad
  I_\Lchiral
  =
  \tfrac{1}{2} ( z - y )
  \ , \quad
  C
  =
  c + \tfrac{1}{2}
  \ \ 
  ( C = R , G , B , \  c = r , g , b )
  \ .
\end{equation}
The electromagnetic charge $Q$ is
\begin{equation}
\label{QIc}
  Q
  =
  \tfrac{1}{2} Y + I_\Lchiral
  =
  z - \tfrac{1}{3} ( r + g + b )
  \ .
\end{equation}
Electroweak symmetry breaking is a loss of $y$-symmetry,
a loss of conservation of $y$-charge.

The following $\Spin(10)$ chart
shows the electron generation of fermions of the standard model
arrayed in columns according to the number of up-bits
(compare Table~4 of \cite{Baez:2009dj}).
The left element of each entry (before the colon) signifies which bits
are up, from -- (no bits up, or $\yd\zd\rd\gd\bd$) in the 0 column,
to $yzrgb$ (all bits up, or $\yu\zu\ru\gu\bu$) in the 5 column;
the right element of each entry is the corresponding fermion,
which comprise (electron) neutrinos $\nu$, electrons $e$,
and up and down quarks $u$ and $d$,
each in right- and left-handed Dirac chiralities $\Rchiral$ and $\Lchiral$,
and each in (unbarred) particle and (barred) anti-particle species,
a total of $2^5 = 32$ fermions:
\begin{equation}
\label{yzrgbtab}
  \begin{array}{c@{\quad\quad}c@{\quad\quad}c@{\quad\quad}c@{\quad\quad}c@{\quad\quad}c}
  \hline
  \multicolumn{6}{c}{\mbox{Fermions and their $\Spin(10)$ bitcodes, arranged by the number of up-bits}}
  \\
  0 & 1 & 2 & 3 & 4 & 5
  \\
  \hline
  \mbox{--} : \
  \bar{\nu}_\Lchiral
  &
  y : \ 
  \bar{\nu}_\Rchiral
  &
  \phantom{d}\bar{c} : \ 
  \bar{u}_\Lchiral^{\bar c}
  &
  \phantom{d}y\bar{c} : \
  \bar{u}_\Rchiral^{\bar c}
  &
  zrgb : \ 
  \nu_\Lchiral
  &
  yzrgb : \ 
  \nu_\Rchiral
\\
  &
  z : \ 
  \bar{e}_\Rchiral
  &
  yz : \ 
  \bar{e}_\Lchiral
  &
  rgb : \ 
  e_\Rchiral
  &
  yrgb : \ 
  e_\Lchiral
  &
\\
  &
  c : \ 
  d_\Rchiral^c
  &
  yc : \ 
  d_\Lchiral^c
  &
  \phantom{d}z\bar{c} : \ 
  \bar{d}_\Rchiral^{\bar{c}}
  &
  \phantom{d}yz\bar{c} : \ 
  \bar{d}_\Lchiral^{\bar{c}}
\\
  &
  &
  zc : \ 
  u_\Lchiral^c
  &
  yzc : \ 
  u_\Rchiral^c
  &
  &
\\
  \hline
  \end{array}
\end{equation}
Here $c$ denotes any of the three colors $r$, $g$, or $b$
(one color bit up),
while ${\bar c}$ denotes any of the three anticolors $gb$, $br$, or $rg$
(two color bits up, the bit flip of a one-color-bit-up spinor).
Every spin group $\Spin(N)$ contains a subgroup $\SU([N/2])$
that preserves the number of up-bits \cite{Atiyah:1964}.
The columns of the chart~(\ref{yzrgbtab})
are $\SU(5)$ multiplets within $\Spin(10)$.
The standard-model group is a subgroup of $\SU(5)$.
All standard-model interactions preserve the number of up-bits.

\section*{Unification of standard-model and spacetime symmetries in the Spin(11,1) geometric algebra}

\noindent
The chart~(\ref{yzrgbtab}) is a riddle of striking features.
The most striking feature is that $\Spin(10)$ chirality
coincides with Dirac chirality.
Chirality counts whether the number of up-bits is even or odd,
with right-handed defined as all bits up.
The odd and even columns of the $\Spin(10)$ chart~(\ref{yzrgbtab})
have respectively right-handed ($\Rchiral$) and left-handed ($\Lchiral$)
Dirac chirality.
Modulo a phase,
chirality is (the eigenvalue of) the pseudoscalar of the algebra,
the
product of all the $N$ vectors in the $N$-dimensional geometric algebra.
The remarkable coincidence of Dirac and $\Spin(10)$ chiralities
suggests that
the vectors of spacetime are related to the vectors of $\Spin(10)$,
in contrast to the usual assumption that the generators
of grand unified symmetries are unrelated to (commute with) those of spacetime.

A second striking feature of the $\Spin(10)$ chart~(\ref{yzrgbtab})
is that each of the $2^5 = 32$ spinors
is itself a 2-component Weyl spinor
(a Dirac spinor of definite chirality, right- or left-handed),
so there are actually $2^6 = 64$ spinors in a generation.
If one asks, what is the smallest geometric algebra that
contains 64 chiral spinors and includes one time dimension,
the answer is the geometric algebra associated with the group $\Spin(11,1)$
of rotations in 11+1 spacetime dimensions.
The reason it is possible to accommodate the 10 dimensions of $\Spin(10)$
and the 4 dimensions of $\Spin(3,1)$ in a spacetime of just 12 dimensions
is precisely that $\Spin(10)$ and $\Spin(3,1)$ redundantly contain the
degrees of freedom associated with flipping chirality.

Adding two extra dimensions to the 10 dimensions of $\Spin(10)$
adds one extra bit, the $t$-bit, or time bit,
to the 5 $yzrgb$ bits of $\Spin(10)$.
The two extra dimensions comprise an eleventh spatial dimension $\bgamma^+_t$
and a time dimension $\bgamma_0 = \im \bgamma^-_t$.

The conventional assumption that $\Spin(10)$ and the Lorentz group $\Spin(3,1)$
combine as a direct product is motivated by
the Coleman-Mandula theorem
\cite{Coleman:1967,Mandula:2015},
which requires that the algebra of any symmetry that has bosonic generators
and yields non-trivial scattering amplitudes
must be a direct product of internal and spacetime algebras.
$\Spin(11,1)$ satisfies the higher-dimensional Coleman-Mandula theorem
\cite{Pelc:1997} trivially, because
the internal and spacetime symmetries of $\Spin(11,1)$ are one and the same.
After grand symmetry breaking,
the Coleman-Mandula theorem requires only that {\em unbroken\/} internal
symmetries commute with those of spacetime.

If indeed $\Spin(10)$ and spacetime symmetries unify in the
$\Spin(11,1)$ geometric algebra,
then the chart~(\ref{yzrgbtab}) cannot be correct as it stands.
The problem is that each of the $2^5 = 32$ spinors in the $\Spin(10)$
chart~(\ref{yzrgbtab}) is a Weyl spinor,
which requires two bits for its description, 
whereas only one extra bit, the $t$-bit, is available.

The key to the riddle of translating
the $\Spin(10)$ chart~(\ref{yzrgbtab})
into $\Spin(11,1)$ is to notice that the spinors in~(\ref{yzrgbtab})
are fermions (unbarred) or antifermions (barred)
as the color chirality $\varkappa_{rgb}$ is positive or negative,
that is, as the number of color up-bits is odd or even.
The five $\Spin(10)$ charges of a spinor are eigenvalues of the five diagonal
bivectors $\bgamma^+_k \bgamma^-_k$, $k = y,z,r,g,b$,
of the geometric algebra.
If these diagonal bivectors are modified by multiplying them
by $\varkappa_{rgb}$, then their eigenvalues will measure the charge
of the fermion, not the antifermion, in all entries of the $\Spin(10)$ chart.
A key point that allows this adjustment to be made consistently
is that $\varkappa_{rgb}$ commutes with all standard-model bivectors.
Notably, $\varkappa_{rgb}$ does not commute
with $\SU(5)$ bivectors that transform between leptons and quarks;
but that is fine,
because $\SU(5)$ is not an unbroken symmetry of the standard model.
A consistent way to implement this modification,
that leaves the bivector algebra of the standard model
(but not of $\SU(5)$)
unchanged,
is to multiply all imaginary bivectors $\bgamma^+_k \bgamma^-_l$
by $\varkappa_{rgb}$,
while leaving all real bivectors
$\bgamma^+_k \bgamma^+_l$ and $\bgamma^-_k \bgamma^-_l$
unchanged,
\begin{equation}
\label{adjustsmbivectorsI6}
  \bgamma^+_k
  \bgamma^-_l
  \rightarrow
  \bgamma^+_k
  \bgamma^-_l
  \varkappa_{rgb}
  \ , \quad
  k,l = t,y,z,r,g,b
  \ .
\end{equation}

The modification~(\ref{adjustsmbivectorsI6})
serves to replace each antifermion in the chart with the corresponding fermion.
For example, the positron entries
$\bar{e}_\Rchiral$ and $\bar{e}_\Lchiral$
are replaced by electrons
$e_\Lchiral$ and $e_\Rchiral$.
What about antifermions?
Where have they gone?
The answer is that antifermions are obtained from fermions in the usual way
\cite{Hamilton:2023a},
by taking their complex conjugates and multiplying by the conjugation operator,
$\bar{\psi} \equiv C \psi^\ast$.
In any geometric algebra with one time dimension,
the conjugation operator $C$ flips all bits except the time bit $t$.
Thus antifermions appear in a second copy of
the $\Spin(10)$ chart~(\ref{yzrgbtab}),
a conjugated version in which all fermions are replaced by antifermions.
The fermionic and antifermionic (conjugated) charts
are distinguished by a flip of the time-bit $t$,
a pretty conclusion.

It requires some work \cite{Hamilton:2022b}
to establish the correct assignment of Dirac boost ($\Uparrow$ or $\Downarrow$)
and spin ($\uparrow$ or $\downarrow$) bits,
but the end result is the following $\Spin(11,1)$ chart of spinors,
arranged in columns by the number of $\Spin(10)$ up-bits
as in the earlier chart~(\ref{yzrgbtab}):
\begin{equation}
\label{tyzrgbtab}
  \begin{array}{c@{\quad}c@{\quad}c@{\quad}c@{\quad}c@{\quad}c}
  \hline
  0 & 1 & 2 & 3 & 4 & 5
  \\
  \hline
  \mbox{--} : \ 
  \substack{\displaystyle
  \bar{\nu}_{\Updown}
  \\\displaystyle
  \nu_{\Downdown}
  }
  &
  y : \ 
  \substack{\displaystyle
  \bar{\nu}_{\Downdown}
  \\\displaystyle
  \nu_{\Updown}
  }
  &
  \phantom{d}\bar{c} : \ 
  \substack{\displaystyle
  \bar{u}_{\Updown}^{\, \bar{c}}
  \\\displaystyle
  u_{\Downdown}^{\, c}
  }
  &
  \phantom{d}y\bar{c} : \
  \substack{\displaystyle
  \bar{u}_{\Downdown}^{\, \bar{c}}
  \\\displaystyle
  u_{\Updown}^{\, c}
  }
  &
  zrgb : \ 
  \substack{\displaystyle
  \nu_{\Downup}
  \\\displaystyle
  \bar{\nu}_{\Upup}
  }
  &
  yzrgb : \ 
  \substack{\displaystyle
  \nu_{\Upup}
  \\\displaystyle
  \bar{\nu}_{\Downup}
  }
\\[1.5ex]
  &
  z : \ 
  \substack{\displaystyle
  \bar{e}_{\Downdown}
  \\\displaystyle
  e_{\Updown}
  }
  &
  yz : \ 
  \substack{\displaystyle
  \bar{e}_{\Updown}
  \\\displaystyle
  e_{\Downdown}
  }
  &
  rgb : \ 
  \substack{\displaystyle
  e_{\Upup}
  \\\displaystyle
  \bar{e}_{\Downup}
  }
  &
  yrgb : \ 
  \substack{\displaystyle
  e_{\Downup}
  \\\displaystyle
  \bar{e}_{\Upup}
  }
  &
\\[1.5ex]
  &
  c : \ 
  \substack{\displaystyle
  d_{\Upup}^{\, c}
  \\\displaystyle
  \bar{d}_{\Downup}^{\, \bar{c}}
  }
  &
  yc : \ 
  \substack{\displaystyle
  d_{\Downup}^{\, c}
  \\\displaystyle
  \bar{d}_{\Upup}^{\, \bar{c}}
  }
  &
  \phantom{d}z\bar{c} : \ 
  \substack{\displaystyle
  \bar{d}_{\Downdown}^{\, \bar{c}}
  \\\displaystyle
  d_{\Updown}^{\, c}
  }
  &
  \phantom{d}yz\bar{c} : \ 
  \substack{\displaystyle
  \bar{d}_{\Updown}^{\, \bar{c}}
  \\\displaystyle
  d_{\Downdown}^{\, c}
  }
\\[1.5ex]
  &
  &
  zc : \ 
  \substack{\displaystyle
  u_{\Downup}^{\, c}
  \\\displaystyle
  \bar{u}_{\Upup}^{\, \bar{c}}
  }
  &
  yzc : \ 
  \substack{\displaystyle
  u_{\Upup}^{\, c}
  \\\displaystyle
  \bar{u}_{\Downup}^{\, \bar{c}}
  }
  &
  &
\\[1.5ex]
  \hline
  \end{array}
\end{equation}
Whereas in the original $\Spin(10)$ chart~(\ref{yzrgbtab})
each entry was a two-component Weyl spinor,
in the $\Spin(11,1)$ chart~(\ref{tyzrgbtab})
the two components of each Weyl spinor appear in bit-flipped entries.
For example, the right-handed electron $e_\Rchiral$ of the original chart
is replaced by $e_{\Upup}$,
and its spatially rotated partner $e_{\Downdown}$ of the same chirality
appears in the all-bit-flipped entry.
Each entry still has two components,
but in the $\Spin(11,1)$ chart those two components differ by their $t$-bit;
the upper component has $t$-bit up,
the lower $t$-bit down.
The net number of degrees of freedom remains the same, $2^6 = 64$.

\spinelevenonechartfig

Figure~\ref{spin111chart} illustrates
one generation (the electron generation) of fermions of the standard model
arranged according to their $\Spin(11,1)$ $tyzrgb$ charges.

The definitive test of the viability of the $\Spin(11,1)$ model
is to write down the relations between the $\Spin(11,1)$ and Dirac
geometric algebras, and to check that the relations satisfy all constraints.
The coincidence of $\Spin(10)$ and Dirac chiralities implies
that the Dirac spacetime vectors must be higher dimensional elements
of the $\Spin(11,1)$ algebra.
The four Dirac spacetime vectors $\bgamma_m$,
$m = 0, 1, 2, 3$
in terms of the algebra of the twelve $\Spin(11,1)$ vectors
$\bgamma_k^\pm$, $k = t,y,z,r,g,b$, are
\cite{Hamilton:2022b}
\begin{equation}
\label{vectorsspin10ew}
  \bgamma_0
  =
  \im \bgamma_t^-
  \ , \quad
  \bgamma_1
  =
  \bgamma_y^- \bgamma_z^- \bgamma_r^+ \bgamma_g^+ \bgamma_b^+
  \ , \quad
  \bgamma_2
  =
  \bgamma_y^- \bgamma_z^- \bgamma_r^- \bgamma_g^- \bgamma_b^-
  \ , \quad
  \bgamma_3
  =
  \bgamma_t^+ \bgamma_y^+ \bgamma_y^- \bgamma_z^+ \bgamma_z^-
  \ .
\end{equation}
The Dirac vectors~(\ref{vectorsspin10ew}) all have grade 1~mod~4.
The multiplication rules for
the vectors $\bgamma_m$ given by equations~(\ref{vectorsspin10ew})
agree with the usual multiplication rules for Dirac $\gamma$-matrices:
the vectors $\bgamma_m$ anticommute,
    and their scalar products form the Minkowski metric.
    All the spacetime vectors $\bgamma_m$ commute with all standard-model bivectors
modified per~(\ref{adjustsmbivectorsI6}).
The Dirac pseudoscalar $I$ coincides with
the $\Spin(11,1)$ pseudoscalar $J$,
\begin{equation}
\label{IJ}
  I
  \equiv
  \bgamma_0 \bgamma_1 \bgamma_2 \bgamma_3
  =
  J
  \equiv
  -\im
  \bgamma^+_t \bgamma^-_t
  \bgamma^+_y \bgamma^-_y \bgamma^+_z \bgamma^-_z
  \bgamma^+_r \bgamma^-_r \bgamma^+_g \bgamma^-_g \bgamma^+_b \bgamma^-_b
  \ .
\end{equation}

Thus the Dirac and standard-model algebras are subalgebras
of the $\Spin(11,1)$ geometric algebra,
such that all Dirac generators commute with all standard-model bivectors
modified per~(\ref{adjustsmbivectorsI6}),
as required by the Coleman-Mandula theorem.

The time dimension $\bgamma_0$
in equations~(\ref{vectorsspin10ew})
is just a simple vector
in the $\Spin(11,1)$ algebra,
but the 3 spatial dimensions $\bgamma_k$, $k = 1,2,3$
are all 5-dimensional.
The spatial dimensions
share a common 2-dimensional factor $\bgamma_y^- \bgamma_z^-$.
Aside from that common factor,
each of the 3 spatial dimensions is itself 3-dimensional:
$\bgamma_r^+ \bgamma_g^+ \bgamma_b^+$,
$\bgamma_r^- \bgamma_g^- \bgamma_b^-$,
and
$\bgamma_t^+ \bgamma_y^+ \bgamma_z^+$.


\section*{The road to grand unification}

\noindent
In the $\Spin(11,1)$ model,
grand unification unifies all four forces,
not just the three forces of the standard model,
and it involves a transition from the 3+1 dimensions of today's spacetime
to higher dimensions.

As long as spacetime is 4-dimensional, as it is today,
any internal symmetry must commute
with the four Dirac vectors~(\ref{vectorsspin10ew}),
in accordance with the Coleman-Mandula theorem.
This is a tight constraint.
There appears to be a unique minimal symmetry-breaking chain
from $\Spin(11,1)$ to the standard model
\cite{Hamilton:2022b},
proceeding by the Pati-Salam group $\Spin(4)_w {\times} \Spin(6)_c$
\cite{Pati:1974},
as first proposed by \cite{Harvey:1980},
and advocated by \cite{Altarelli:2013b,Babu:2015},
\begin{align}
\label{symbreaking}
  &
  \Spin(11,1)
  \ \underset{??}{\parbox{2em}{\rightarrowfill}} \
  \Spin(10,1)
  \ \underset{10^{15} \unit{GeV}}{\parbox{3em}{\rightarrowfill}} \
  \Spin(4)_w \times \Spin(6)_c \times \Spin(3,1)
  \ \underset{10^{12} \unit{GeV}}{\parbox{3em}{\rightarrowfill}} \
\nonumber
\\
  &\
  \U(1)_Y \times \SU(2)_\Lchiral \times \SU(3)_c \times \Spin(3,1)
  \ \underset{160 \unit{GeV}}{\parbox{3em}{\rightarrowfill}} \
  \U(1)_Q \times \SU(3)_c \times \Spin(3,1)
  \ .
\end{align}
The top line of the chain~(\ref{symbreaking}) is the prediction,
while the bottom line is the standard model.
Note that
the grand unified group is $\Spin(10,1)$ rather than $\Spin(11,1)$ itself.
%
%
The predicted energy scales of unification are deduced from the
running of the three coupling parameters
of the standard model.

The minimal Higgs sector that mediates symmetry breaking is likewise unique,
consisting of the dimension 66 bivector (adjoint) representation of Spin(11, 1).
In effect,
the Lorentz-scalar (spin~0) Higgs sector
matches the Lorentz-vector (spin~1) gauge sector.
The general principles underlying symmetry breaking by the Higgs mechanism
\cite{Englert:1964,Higgs:1964}
are:
(1) the Higgs field before symmetry breaking is a scalar (spin~0) multiplet
of the unbroken symmetry;
(2) one component of the Higgs multiplet acquires a nonzero vacuum
expectation value;
(3) components of the Higgs multiplet whose symmetry is broken
are absorbed into longitudinal components of the broken gauge (spin~1) fields
by the Goldstone mechanism \cite{Goldstone:1962},
giving those gauge fields mass;
and (4) unbroken components of the Higgs multiplet persist as scalar fields,
potentially available to mediate the next level of symmetry breaking.

The 66-component Higgs multiplet
contains a four-component multiplet with generators
$\bgamma^+_t \bgamma^\pm_k$, $k = y,z$
(adjusted per~(\ref{adjustsmbivectorsI6})),
whose properties match those of the electroweak Higgs multiplet
required by the standard Weinberg \cite{Weinberg:1967}
model of electroweak symmetry breaking.
The Higgs multiplet breaks electroweak symmetry
when it acquires a vacuum expectation value
$\langle \bH \rangle$
proportional to
$\bgamma^+_t \bgamma^-_y$,
\begin{equation}
\label{Higgsvac}
  \langle \bH \rangle
  =
  \langle H \rangle
  \bgamma^+_t \bgamma^-_y
  \varkappa_{rgb}
  \ .
\end{equation}
The factor of the color chiral operator $\varkappa_{rgb}$
is from the adjustment~(\ref{adjustsmbivectorsI6}).
The electroweak Higgs field $\langle \bH \rangle$
breaks $y$-symmetry,
carries one unit of $y$ charge,
and gives masses to fermions by flipping their $y$-bit.

The grand Higgs field
$\langle \bT \rangle$
that breaks $\Spin(10,1)$
to the Pati-Salam group
is proportional to the time bivector
$\bgamma^+_t \bgamma^-_t$,
\begin{equation}
\label{Higgsnuvac}
  \langle \bT \rangle
  =
  -\im
  \langle T \rangle
  \bgamma^+_t \bgamma^-_t
  \varkappa_{rgb}
  \ .
\end{equation}
Again, the factor of the color chiral operator $\varkappa_{rgb}$
is from the adjustment~(\ref{adjustsmbivectorsI6}).
The grand Higgs field $\langle \bT \rangle$ generates a Majorana mass term
for the right-handed neutrino by flipping its $t$-bit.
Only the right-handed neutrino can acquire a Majorana mass,
because only the right-handed neutrino has zero standard-model charge;
for other fermions, flipping the $t$-bit is prohibited by
conservation of standard-model charge.
As is well known, a large Majorana mass, coupled with a smaller Dirac mass,
can explain the small mass of left-handed neutrinos,
by the see-saw mechanism \cite{GellMann:1979}.

A leading idea of the standard model of cosmology is that inflation
in the early universe was driven by the energy associated with
grand unification,
e.g.~\cite{Martin:2013,Kumar:2018}.
The grand Higgs field
$\langle \bT \rangle$
is available to drive cosmological inflation.

The bivectors
$\bgamma^+_t \bgamma^\pm_k$, $k = y,z$
generate the electroweak Higgs field,
but they cannot generate a gauge symmetry,
because they fail to commute with the
Higgs field $\langle \bT \rangle$ that breaks grand symmetry,
equation~(\ref{Higgsnuvac}).
The only apparent solution to this problem
is to postulate that the dimension $\bgamma^+_t$ is a scalar dimension
that does not generate any symmetry,
a possibility discussed in \S4.4 of \cite{Hamilton:2023a}.
The dimension $\bgamma^+_t$ stands out
as the only spacelike vector of $\Spin(11,1)$
that is not a factor in any unbroken gauge symmetry of the standard model.
The compactification of some dimensions is a well-known feature of
string theories \cite{Becker:2007,Freedman:2012},
but how $\Spin(11,1)$ might break to $\Spin(10,1)$ is
a matter for future research.

\section*{String theory?}

It can scarcely escape notice that $\Spin(10,1)$
has the same number 11 of spacetime dimensions as maximal supergravity
\cite{Freedman:2012,vanNieuwenhuizen:2004rh,Ferrara:2017hed,Deser:2018},
the low-energy limit of M theory \cite{Schwarz:1999,Becker:2007},
which is the conjectured extension of string theory
to include higher-dimensional objects, branes.
Extensions of string theory to 12 spacetime dimensions, F-theory,
have also been conjectured \cite{Vafa:1996,Heckman:2010,Callaghan:2012}.
String-theory-inspired models usually assume that
the parent spacetime is, at least locally,
a product space, consisting of 4 large dimensions
multiplied by a space of compactified or hidden dimensions.
By contrast,
in the $\Spin(11,1)$ geometric algebra,
although the time dimension is a vector,
each of the 3 spatial Dirac dimensions
is a pentavector, a 5-dimensional multivector,
equations~(\ref{vectorsspin10ew}).
The spatial dimensions share a common 2-dimensional factor,
and beyond that are each 3-dimensional.
Is this arrangement viable in string theory?

\section*{Conclusion}

\noindent
If the ideas in this essay are correct,
then the DNA of the Universe is written in a language whose letters
are the six bits of spinors in 11+1 spacetime dimensions.
The $\Spin(11,1)$ geometric algebra describes only the tangent
space of the spacetime,
not the geometry of the 11+1~dimensional spacetime itself.
One may speculate, as string theorists have done,
that the complexity of the laws of physics is associated with
the complicated geometry of the hidden extra dimensions
and the fields that wrap them.
Like the DNA of life, the letters may be simple,
but the stories written with those letters could be fabulously complex.

This essay is based on original research presented by \cite{Hamilton:2022b}.


\printbibliography

\end{document}